# Screening and Fabrication of Half-Heusler phases for thermoelectric applications

Wilfried Wunderlich[1], Yuichiro Motoyama[1]
1) Tokai University, Fac. Eng, Materials Science Dept., Hiratsuka-shi, Japan

**ABSTRACT**

Half-Heusler phases have gained recently much interest as thermoelectric materials. Screening of possible systems was performed by drawing their stability region in a three-dimensional Pettifor map. The fabrication of Half-Heusler phases requires three steps, surface activation of the raw material by ball milling, arc-melting of pressed pellets and finally long-term annealing treatment in a vacuum furnace. On doped TiCoSb specimens, Seebeck coefficients of 0.1 mV/K, on NiNbSn 0.16 mV/K were measured, although the microstructure was not yet optimized.

**INTRODUCTION**

Thermoelectric materials (TE) are considered as clean energy sources helping to solve the severe $CO_2$- problem, but materials with higher efficiency need to be found. The figure-of-merit $ZT=S^2\sigma T/\kappa$ requires a high Seebeck coefficient $S$ and electric conductivity $\sigma$ and low thermal conductivity $\kappa$. For increasing ZT several concepts for materials design of thermoelectrics have been introduced [1-2], such as phonon-glass, electron-crystal, (PGEC), heavy rattling atoms as phonon absorbers, high density of states at the Fermi energy, differential temperature dependence of density of states, high effective electron mass [3], superlattice structures with their confined two-dimensional electron gas [4] and electron-phonon coupling [5,6]. In this study we focus on the search and fabrication of Half-Heusler (HH) structures, which have been found as successful thermoelectric materials, like NiTiSn [7,8]. The reason why HH [7-9], perovskite [3-5] and Skutterudite [2] are successful is sketched in fig. 1. The phonon wave pushes the electron waves through the crystal, when the electron-phonon coupling has suitable interaction energy [6]. This can be successful, when the electron waves have enough freedom to vibrate. These three crystal structures have vacant lattice positions or force atoms to sit in larger atomic distances than according to their atomic spheres. The empty space is one of the necessities for good thermoelectric materials.

In the first section the search for new Half-Heusler phases by three-dimensional Pettifor maps is described. The second and third sections describe the fabrication, experimentally obtained microstructures and the thermoelectric properties of different systems which were selected because they are possible candidates for HH alloys according to the 3- Pettifor maps.

**THREE-DIMENSIONAL PETTIFOR MAPS**

Drawing of Pettifor maps, see e.g. [10, 11] is the suitable method to display regions of element combinations, in which certain crystal structures are stable. To our best knowledge it is the first time to show such a map for ternary components XYZ. In HH-phases elements on each position of the three positions $X_1$ Y Z come from different groups of the periodic table (fig. 2).

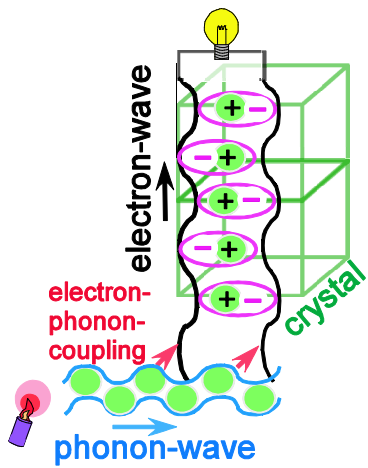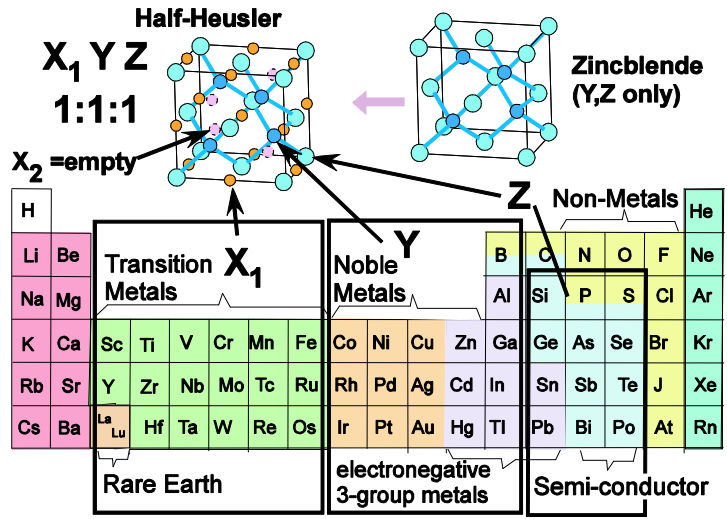

**Figure 1.** Sketch of the inter-action between phonon- and electron-waves in proper thermoelectric crystals

**Figure 2.** Half-Heusler crystal structure with $X_1$ Y Z positions occupied with elements from different groups in the periodic table. If $X_1$ is empty, Zincblende structure is obtained. If $X_2$ is occupied, Full-Heusler is obtained.

HH-phases are related to Full-Heusler (FH) phases $X_1X_2YZ$ by leaving position $X_2$ empty and related to the Zincblende structure leaving position $X_1$ also empty (fig. 2). The crystallographic

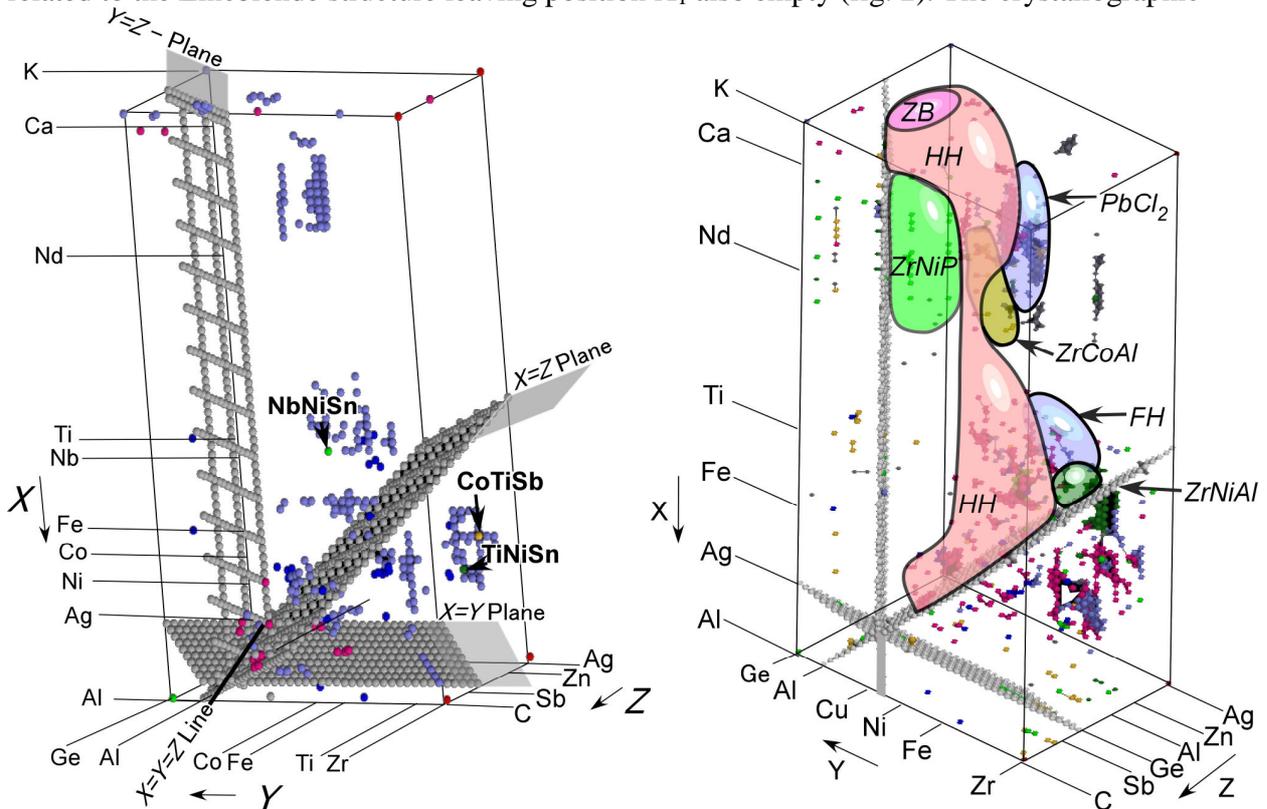

**Figure 3.** Three-dimensional Pettifor maps of XYZ compounds according to [12]; a) Half-Heusler structures only, b) regions of different crystal structures as marked (HH= Half-Heusler, FH=Full-Heusler, $PbCl_2$, ZrNiP, ZrCoAl, ZrNiAl).

**Table 1.** Half-Heusler structure and their competing crystal phases with their space group, reference name, a typical representative, and the number of representatives shown in fig. 3b.

| Crystal structure | Space group (SG) # | referred in literature as | Representative | appeared in Fig. 3 b n-times |
|---|---|---|---|---|
| HH | F -4 3m (216) | MgAgAs | TiCoSb | 111 |
| FH | F m -3 m (225) | VFe$_2$Si | TiNi2Al | 26 |
| ZB | F d -m 3 (216) | Zincblende | GaInSb | 18 |
| TiNiSi | P n a m (062) | TiNiSi (PbCl$_2$) | CrNiSb | 97 |
| ZrNiAl | P -6_2 m (189) | ordered Fe$_2$P | CrNiAs | 53 |
| FeSiV | P6/mmm (191) | CeNiSb | FeSiV | 53 |
| ZrNiP | P6_3/mmc (194) | ZrBeSi (MgZn$_2$) | ZrNiP | 63 |
| ZrCoAl | F d-3m S (227) | CoMnSb (CaF$_2$) | ZrCoAl | 80 |
| fcc-SS. | F m 3 m (225) | Ni | AgAuSi | – |
| bcc-SS. | I m 3 m (229) | Nb | | – |
| hcp-SS | P6_3/mmc (194) | Cr | | – |
| others | | | | 89 |

*) also available as anti-PbCl$_2$ structure; atomic positions strongly depend on the ionic radii.
Aberrations: HH=Half-Heusler, FH=Full-Heusler, ZB=Zincblende, SS=Solid Solution.

database [12] was searched for XYZ components and their crystallographic structures were put in the positions of sets of large tables, where x-, y- and z- axis correspond to each element on the crystallographic positions $X_1$, Y, Z. In Pettifor maps the elements are ordered according to the Mendeleev number [11], which starts with 1 (He), and ends at 103 (H). The regions for elements on X, Y, Z were restricted (fig. 3) because of the table's size. Symmetry planes with equal X=Y, Z=X and X=Y-elements are marked in fig. 3 with grey dots. Most of the crystal structure of XYZ compounds in the mentioned element range consists of those phases shown in table 1: HH-structures, in literature referred as MgAgAs, as shown in the third row together with the other XYZ-phases. These eight crystal structures cover more than 80% of all combinations in the 3-d Pettifor map (fig. 3b) and with ab-initio calculations [13] their stability and bond spectrum [14] was checked. HH structures cover a region much larger than from fig, 2 expected, limited by InLiAg, SbLiAg, PdFeTe and CuAgTe at each end, leaving many topics left for new research.

**EXPERIMENTAL: FABRICATION AND MICROSTRUCTURE**

Six alloy systems were selected in order to investigate whether HH phase exists as a stable phase. The metallic powders (μm-size, purity 99.999%, Fine Chemicals, Japan) with defined weight ratios were ball-milled for about 4h with 5mm Zirconia balls and then pressed into 8mm sized pellets. Subsequent 10kW arc-melting still leaves non-reacted Ti as black inclusions (fig. 4a) at TiCoSb, but during sintering of powder under Ar at 1023K for 10h 50μm sized HH-crystals were formed (fig. 5a) also confirmed by XRD diffraction, details see [15]. In the case of Mo its high melting point prevents sufficient diffusion but annealing under Ar at 1023K for 100h reduced the Mo particles remarkably from 15 μm to about 5 μm observed by SEM (Hitachi S-3200N, 30kV) equipped with EDS (Noran, Be-window). The microstructure after arc-melting is shown in fig. 4, a) TiCoSb, b) ZrCoSb, c) MoFeSb, d) NiCrSn, e) NbNiSn, f) NiTiSn. In all systems, the non-reacted transition metal elements Ti, Cr, Zr, Nb appear darker compared to Ni, Sn, or Sb. During solidification Sb (upper row) forms an intermetallic compound with Co or Fe and Sn (lower row) with Ni forming $Ni_3Sn_2$, $Ni_3Sn_4$ or NbNiSn. After annealing at 1023 K for 100h (fig. 5) diffusion drives microstructure homogeneous and binary particles smaller. In the TiCoSb system except the HH phase no other ternary phase was observed, in agreement with a study [16] on ball-milling, which aims a surface activation and hence higher reactivity. However, with increasing milling time, separation into two phases Ti and CoSb occurred, in agreement with our results on planar interface geometry [15]. In the case of ZrCoSb, HH is dominating, but also an eutectic $CoSb_3$-Zr microstructure remains due to non-stoichiometry. The MoFeSb system shows besides HH the $Mo_2FeSb$ full-Heusler phase. According to the Pettifor map (fig. 3b) the CrNiSn system lies within the range of possible HH phases but it showed about 5μm large cubic-shaped $Cr_2NiSn$ crystals, confirmed as full-Heusler phase. In the NbNiSn system after annealing hardly discernible NbNiSn, $Nb_3Sn_2$, $Nb_3Sn_4$ or $Nb_3Ni_2Sn_2$ phases appear. For the NiTiSn system decomposition into $Ni_3Sn_4$ and Ti occurred, indicating that the annealing temperature 1023K was higher than the temperature for which NiTiSn HH-phase is stable. Table 2 summarizes the results and indicates necessary optimization.

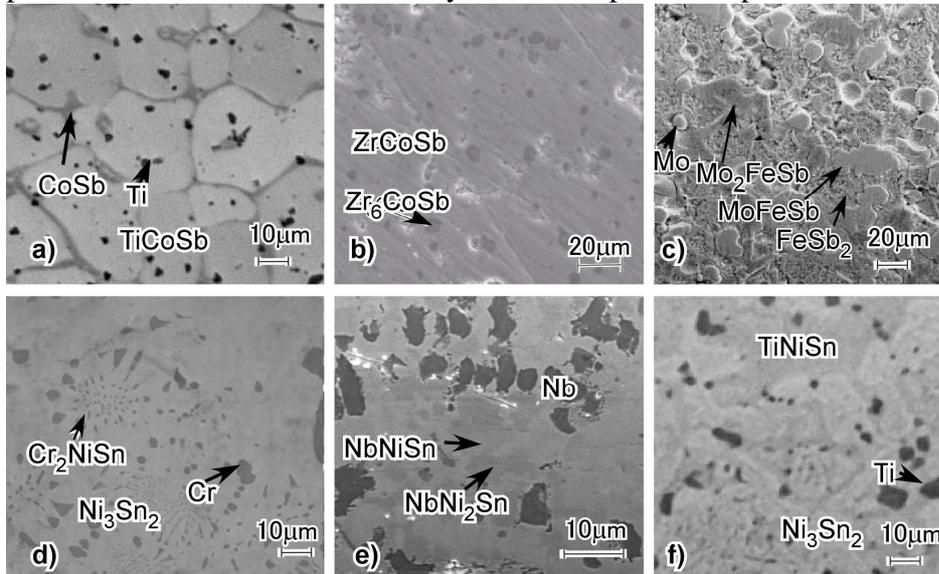

**Figure 4.** SEM micrographs of arc-melted specimens, showing the phases as marked; a) TiCoSb (HH), grey CoSb, dark Ti b) ZrCoSb, dark $Zr_6CoSb$, c) MoFeSb, dark Mo, bright FeSb, d) $Cr_2NiSn$, dark Cr, bright $Ni_3Sn_2$; e) NbNiSn, black Nb, dark $NbNi_2Sn$, bright NbNiSn or $Nb_3Ni_2Sn_2$, f) NiTiSn, dark Ti, bright $Ni_3Sn_2$.

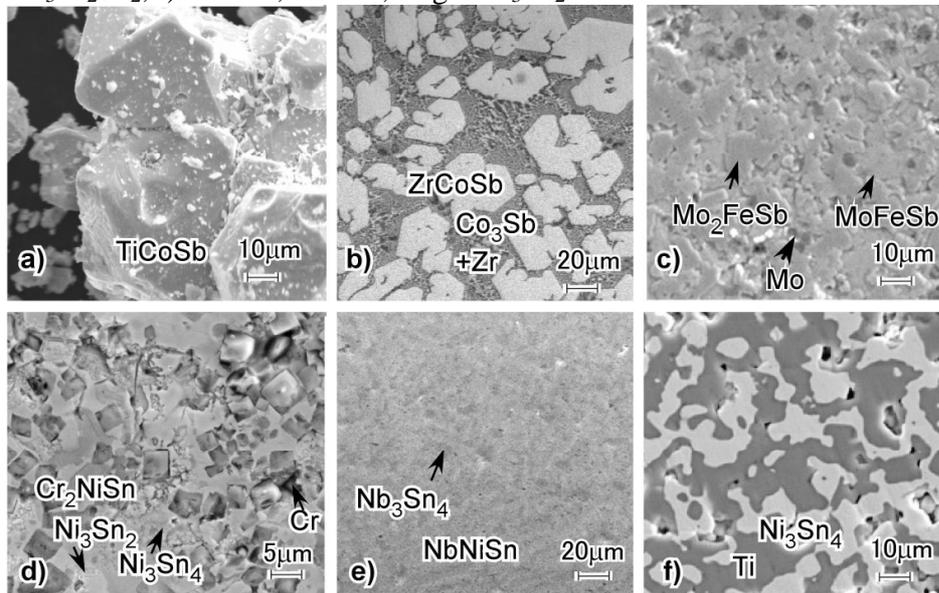

**Figure 5.** SEM micrographs after annealing (1023K 100h) of the phases, a) TiCoSb HH-crystals, b) ZrCoSb, dark Zr, bright CoSb, c) MoFeSb, dark Mo, grey $Mo_2FeSb$, d) $Cr_2NiSn$-Full-Heusler crystals, dark Cr, bright $Nb_3Sn_2$, $Nb_3Sn_4$; e) NbNiSn, dark $Nb_3Sn_2$, $Nb_3Sn_4$, or $Nb_3Ni_2Sn_2$, f) dark Ti and bright $Ni_3Sn_4$.

**EXPERIMENTAL: SEEBECK- VOLTAGE MEASUREMENTS**

The Seebeck voltage was measured in a self-built device made of refractories as explained in [17]. The pellet-shaped specimen lies with one end on a 15x15x2mm sized ceramics heater (Sakaguchi MS-1000) heated up to 1273K and the other end on a 30x10x10mm sized Cu heat sink. The Seebeck voltage $U_S$ was measured with 1mm thick Ni-wires lying below and above the specimen with attached voltmeters [17] and were recorded together with temperature data in-situ in an attached computer. While Seebeck voltage measurements reported in literature are usually measured with small temperature gradients, for which theoretical conditions are valid for, in our case large temperature gradients of more than 600K are present. Such measurement conditions lead to somewhat higher values for the Seebeck voltages as literature data [9,17]. When heat flows steadily, the electric circuit of hot and cold end of the specimen was connected via resistances of $R$=1Ω, 100Ω, 10kΩ, or 1MΩ under closed circuit condition, and the electric current was measured and recorded.

The second and third row of table 2 show crystal structure calculation results explained elsewhere [14], the forth row summarizes the microstructure and the last two rows show Seebeck voltage at ΔT=600K and its maximum electric current. The maximum Seebeck voltage was measured in our apparatus on doped $NaTaO_3$ with Seebeck voltage -250mV and a maximum current of -330 μA [17], but the HH-specimens presented here, showed much smaller values. NbNiSn alloys show typically $U$=-100mV and $I$=-11mA, which yield to a Seebeck-coefficient $S$=-100/600 mV/K=-0.16 mV/K. The Seebeck voltage strongly depends on the stochiometry and was highest for alloys with $Nb_{10}Ni_{10}Sn_8$ concentration. TiNiSn doped with 3at% Fe showed a positive Seebeck-voltage of 58mV. To conclude, further increase in Seebeck voltage is expected due to doping and microstructure optimization.

**Table 2** Summary of calculation and experimental results

| System | Calculation Results | | Experimental Results | | |
|---|---|---|---|---|---|
| | Stable Lattice in XYZ order | Lattice constant $a$ [nm] | Microstructure after annealing at 1023K for 100h | Seebeck-Voltage at ΔT=600K | |
| | | | | $U$ [mV] | $I_{max}$ [μA] |
| TiCoSb | TiCoSb | 0.5873 | HH; Ti+CoSb +3%Fe | −70 | −3 |
| ZrCoSb | ZrCoSb | 0.5702 | ZrCoSb, $Zr_6CoSb$ | 3 | 30 |
| MoFeSb | MoFeSb | 0.5588 | Mo, MoSbFe, FeSb | 7 | 11 |
| NiCrSn | $Cr_2NiSn$ | 0.5902 | $Cr_2NiSn$, Cr, $Ni_2Sn$ | 7 | 12 |
| NbNiSn | NbNiSn | 0.5987 | HH, $Nb_5Ni_2Sn_2$, $Ni_3Sn_2$ | −100 | −11 |
| TiNiSn | TiNiSn | 0.5919 | NiSnTi, $NiTi_3Sn$, $NiTiSn_3$ +3at%Fe | 58 | 0.6 |

**CONCLUSIONS**

1) Half Heusler (HH) phases are promising candidates for thermoelectric materials approached from the metallic side due to one vacant lattice position. They are found to be stable phase over a wide range of element combinations as displayed here for the first time as three-dimensional Pettifor maps.
2) The best fabrication method for HH found in this study was, first ball-milling in order to activate the raw powder's surface, then arc melting and annealing.
3) Seebeck voltage measurements under large temperature gradient are able to judge the performance of thermoelectric candidate materials. Different behavior was found for some alloys when the voltage-current dependence was measured under closed electric circuit conditions, suggesting that the electric current measurements should also be included when comparing the performance of thermoelectrics.

**ACKNOWLEDGEMENTS**

Obtaining experimental data is acknowledged to Yoshikaji Aoki, Kosuke Nakatsuka, Kouta Okayama, Takayuki Nakagome, Kenji Uematsu, Souichiro Yoshimura.